\documentclass[pra,preprintnumbers,amsmath,amssymb,letterpaper,twocolumn]{revtex4-1}

\usepackage{graphicx}
\usepackage{dcolumn}
\usepackage{bm}
\usepackage{epstopdf}
\usepackage{latexsym}
\usepackage{epsfig}
\usepackage{verbatim}
\usepackage{hyperref}
\usepackage{color}

\newcommand{\be}{\begin{equation}}
\newcommand{\ee}{\end{equation}}
\newcommand{\bea}{\begin{eqnarray}}
\newcommand{\eea}{\end{eqnarray}}

\begin{document}


\title{Critical Field Behavior of a Multiply-connected Superconductor \\ in a Tilted Magnetic Field} 

\author{F.N. Womack and P.W. Adams}
\affiliation{Department of Physics and Astronomy, Louisiana State University, Baton Rouge, Louisiana 70803, USA}
\author{J.M. Valles}
\affiliation{Department of Physics, Brown University, Providence, RI 02912, USA}
\author{G. Catelani}
\affiliation{JARA Institute for Quantum Information (PGI-11), Forschungszentrum J\"ulich, 52425 J\"ulich, Germany}

\date{\today}

\begin{abstract}
We report magnetotransport measurements of the critical field behavior of thin Al films deposited onto multiply connected substrates.  The substrates were fabricated via a standard electrochemical process that produced a triangular array of 66 nm diameter holes having a lattice constant of 100 nm.  The critical field transition of the Al films was measured near $T_c$ as a function of field orientation relative to the substrate normal.  With the field oriented along the normal ($\theta=0$), we observe reentrant superconductivity at a characteristic matching field $H_m=0.22$ T, corresponding to one flux quantum per hole.  In tilted fields, the position $H^*$ of the reentrance feature increases as $\sec(\theta)$, but the resistivity traces are somewhat more complex than those of a continuous superconducting film.  We show that when the tilt angle is tuned such that $H^*$ is of the order of the upper critical field $H_c$, the entire critical region is dominated by the enhanced dissipation associated with a sub-matching perpendicular component of the applied field.  At higher tilt angles a local maximum in the critical field is observed when the perpendicular component of the field is equal to the matching field.
\end{abstract}

\maketitle

\section{Introduction}

Spatial confinement can provide a powerful probe of the underlying quantum properties of condensed matter systems.  This is particularly true in superconducting systems for which a variety of confinement strategies have led to the discovery of a diverse range of quantum behavior, including the Josephson effect \cite{Tinkham}, Little-Parks oscillations \cite{LittleParks,YLiu}, Zeeman-limited superconductivity \cite{Catelani,Adams1}, and the even-odd parity asymmetry in superconducting grains \cite{Ralph,Black}.  Recently there has been a renewed interest in using a multiply connected geometry to explore the quantum insulator-to-superconductor transition \cite{Hebard,Shahar,Adams2,Steiner,Goldman1} in homogeneously disordered BCS superconducting films.  In particular, films deposited onto porous substrates can, under the right conditions, exhibit flux quantization effects that reflect the local phase coherence of the superconducting ground state \cite{Welp,Higgins}.  This strategy was used to show that superconducting pair correlations can exist well into the insulating phase of highly disordered Bi films \cite{Valles1,Valles2,Valles3}.  Here we present a study of flux quantization effects in a relatively low disorder spin-singlet BCS superconductor.   We have performed transport measurements on thin Al films deposited onto anodized aluminum oxide substrates patterned with a nano-honeycomb array of holes.  In perpendicular field the magnetoresistance curves exhibit a well-defined reentrant feature when the condition of one flux quantum per substrate hole is achieved.  In order to tune the position of the reentrance we performed critical field measurements in tilted fields. In contrast to data from uniform films, the family of tilted-field traces exhibit multiple crossings and a variety of non-monotonic behaviors.

\section{Sample preparation}

Multiply connected superconducting films were formed by depositing a thin Al layer onto nano-perforated anodic aluminum oxide (AAO) substrates.  The AAO substrates consisted of a triangular array of 66 nm diameter holes.  The lattice constant of the array was 100 nm and the narrowest portion of the superconducting necks between adjacent holes was $\sim30$ nm in width. Details of the preparation and characterization of the AAO substrates have been published elsewhere \cite{Valles1}.  The Al films were formed by e-beam deposition of 99.999\% Al onto AAO substrates held at 84 K. The depositions were made in a typical vacuum $P<3\times10^{-7}$ Torr at a rate of $\sim0.2$ nm/s. Films with thicknesses ranging from 6 to 9 nm had normal state sheet resistances that ranged from $R=300$ to $800\,\Omega$ at low temperature. Magnetotransport measurements were made on a Quantum Design Physical Properties Measurement System via a horizontal rotator insert.  The maximum applied field was 9 T and the base temperature of the system was 1.83 K.  The resistivity measurements were carried out using a standard 4-wire technique.

\section{Experimental results}

In general the critical field of a thin film superconductor has both an orbital and a Zeeman component.  The latter originates from the Zeeman splitting of the conduction electrons.   In most circumstances, however, the orbital response of the superconductor dominates its critical field behavior in the sense that the Zeeman critical field can be an order of magnitude larger that its orbital counterpart.  This is particularly true in high spin-orbit scattering superconductors such as Nb and Pb due to the fact that even relatively modest spin-orbit scattering rates can dramatically quench the Zeeman response \cite{SO}.  But if one makes a low atomic mass film, such as Al, sufficiently thin and orients the field parallel to the film surface then the orbital response will be suppressed and one can realize a purely Zeeman-mediated critical field transition \cite{Adams3}.  In this series of experiments we have explored the critical field behavior of a low spin-orbit, multiply connected, superconductor under conditions in which the orbital and Zeeman contributions to the critical field transition are comparable.  In practice, this can be accomplished by performing the measurements in a tilted magnetic field.

\begin{figure}
\begin{flushleft}
\includegraphics[width=.44\textwidth]{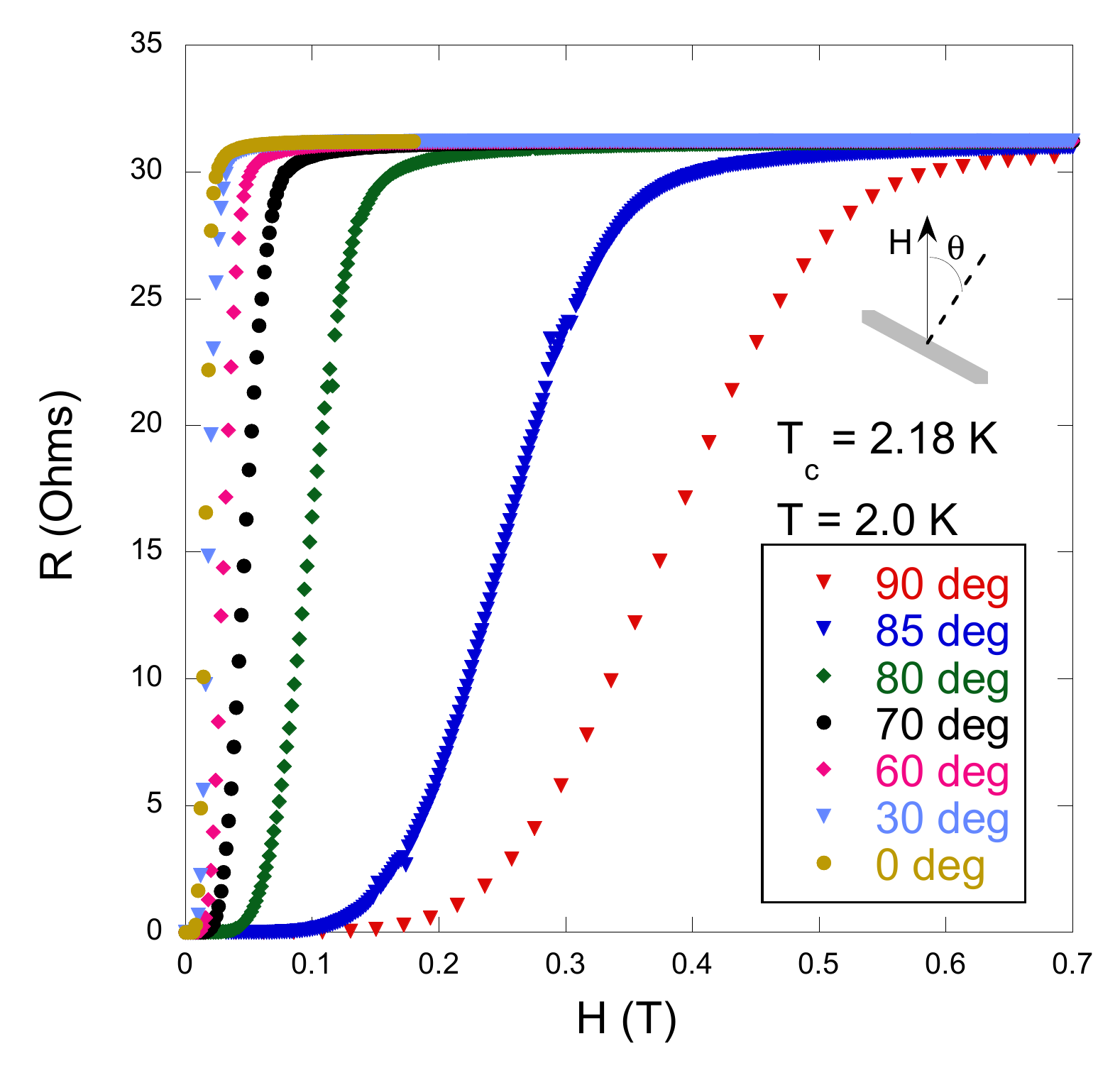}\end{flushleft}
\caption{Plot of the critical field transition of a 9 nm-thick Al film on glass as a function of the angle between the applied field and the normal to the film surface. }
\label{R-H-uniform}
\end{figure}

In addition to producing pair-breaking, the applied magnetic field can also induce flux quantization effects due to the presence of the nano-pore array.  Most of our data was taken in the limit in which the superconducting vortex core radius was comparable to, or larger, than the intra-pore neck widths.  In this limit the vortex cores cannot reside in the neck regions and are therefore relegated to the pores.  This condition can be achieved by exploiting the temperature dependence of the superconducting coherence length \cite{Tinkham},

\be
\xi(T)=0.855\left[\frac{\xi_0\ell}{1-t}\right]^{\frac{1}{2}},
\label{coherence}
\ee
where $\xi_0$ is the BCS coherence length, $\ell$ is the mean-free-path, and $t={T}/{T_c}$.  Since the coherence length grows rapidly as the transition temperature is approached, one can perform the critical field measurements very close to the transition temperature with the vortex cores restricted to the interior of the pores.   Under these conditions the flux quantization effects are maximized.

Before addressing the critical field behavior of the nano-pore Al films, it is useful to establish the behavior of a uniform Al film of similar thickness.  Shown in Fig.~\ref{R-H-uniform} is a plot of tilted-field transitions of a uniform, 9~nm-thick,  superconducting Al film deposited on fire-polished glass.  The superconducting state responds to the applied field in three primary ways.  The first is the formation of quantized vorticity which is entirely associated with the perpendicular component of the field, $H_\perp$.  The second is an orbital pair breaking effect arising from the parallel component of the applied field and the third is pair breaking arising from Zeeman splitting $E_z=g\mu_BH$ of the conduction electrons, where $\mu_B$ is the Bohr magneton, and $g$ is the Land\'e g-factor.  Note that the vortex-mediated perpendicular ($\theta=0$) critical field, usually denoted as $H_{c2}$, is much smaller than its parallel counterpart.  Indeed, from the data in Fig.~\ref{R-H-uniform} we find ${H_{c\parallel}}/{H_{c2}}\sim30$!  From previous studies of Zeeman limited superconductivity in Al films~\cite{CWA}, we estimate that the orbital and the Zeeman contributions to the parallel critical field are comparable in a 9~nm thick film.  In films with thicknesses lower than $\sim4\,$nm, the parallel critical field is completely dominated by the Zeeman term.  Indeed, at low temperatures the purely Zeeman-mediated critical field transition in Al films is first-order at $H_{c\parallel}\sim5\,$T.

\begin{figure}
\begin{flushleft}
\includegraphics[width=.44\textwidth]{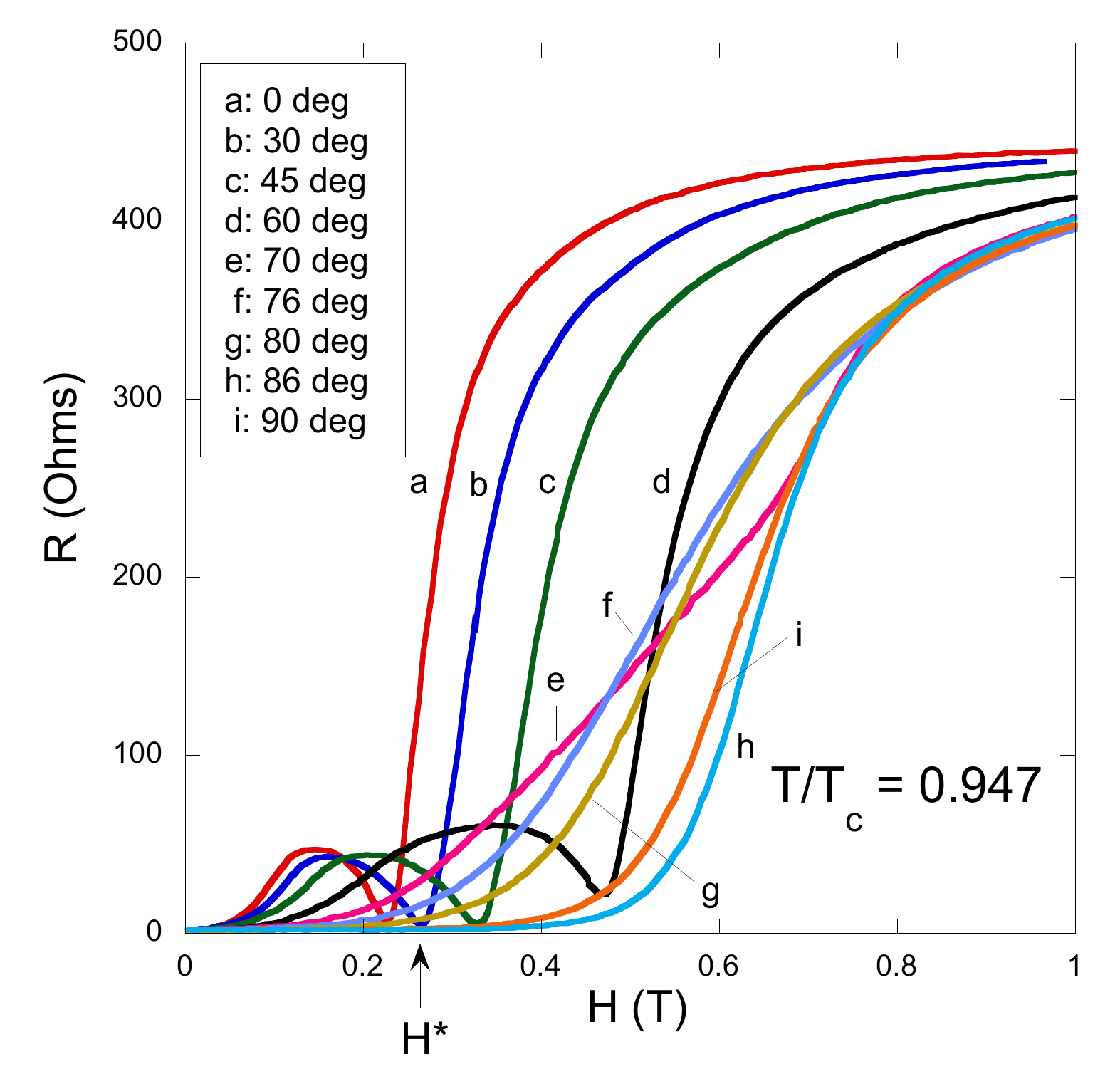}\end{flushleft}
\caption{Resistive critical field transitions of a 9 nm-thick Al film on a nano-honeycomb substrate near $T_c = 2.27$ K.  The dips in the traces occur when the perpendicular component of the applied field equals the matching field $H_m=0.22$ T.  The dip location is a function of angle and is denoted by $H^*$. The figure legend represents the angle between the magnetic field and the normal to the face of the substrate.}
\label{R-H-9}
\end{figure}

In Fig.~\ref{R-H-9} we present critical field data of a 9 nm Al film on an AAO substrate at the reduced temperature $t=0.947$.  Using Eq.~\ref{coherence} we estimate the coherence length $\xi\sim140$ nm, which is much larger than both the radius of the pores and the width of the superconducting necks. Although the $T_c$ of this film was similar to that of its uniform counterpart in Fig.~\ref{R-H-uniform}, its sheet resistance was approximately an order of magnitude higher.  This, of course, is expected due to the fact that the nano-pore film has a relatively small area of metallic coverage. The AAO substrate produces a network of superconducting necks.  The neck regions between the pores are approximately 30 nm in width.  We estimate that the nominal resistance of an AAO film is approximately a factor of 3 larger than its uniform counterpart due to the geometry of the substrate.  In addition, we believe that the resistivity of the neck regions is substantially higher than that of a uniform film of the same thickness due to the surface roughness of the AAO substrates \cite{Valles3}.

\begin{figure}
\begin{flushleft}
\includegraphics[width=.44\textwidth]{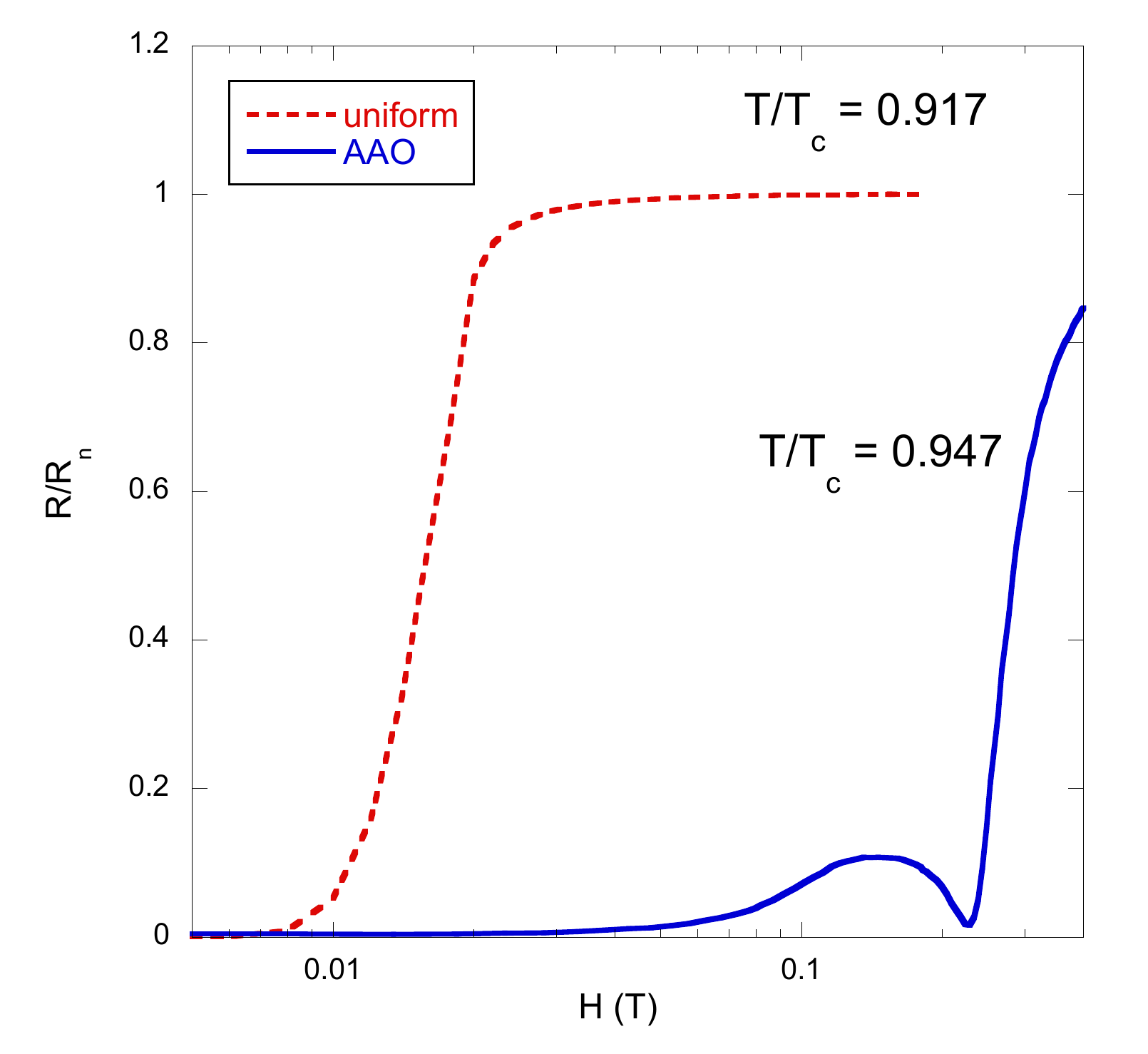}\end{flushleft}
\caption{Resistance as a function of perpendicular field of a 9 nm-thick Al film on glass and a  9 nm-thick Al film on an AAO substrate.}
\label{R-H-lowfield}
\end{figure}

Clearly, the angular dependence of the nano-pore critical field traces is much more complex than that of the uniform film.  We begin by considering the perpendicular field trace ($\theta=0$).  As the field is increased from zero, the resistance rises due to the pair-breaking effects of macroscopic screening currents flowing through the intra-pore superconducting network.  Interestingly, the low-field pair-breaking dissipation is much smaller in the nano-pore film than it is in the corresponding uniform film, as is evident in Fig.~\ref{R-H-lowfield}, indicating a more resilient superconducting state in the AAO film.  This is most likely due to a combination of the finite width and higher resistivity of intra-pore links in the superconducting network.  In fact, the orbital pair-breaking energy is proportional to $Dd^2$, where $D$ is the conduction electron diffusivity and $d$ is the lateral dimension.  As the field is increased, superconducting vorticity moves into the array with the vortex cores residing in the pores.  This produces flux quantization effects that are superimposed onto the orbital pair-breaking background \cite{Crabtree}. When the condition of one vortex per unit cell is reached, there is a net cancellation of the screening currents and a corresponding dip is observed in the R-H trace \cite{Patel}.   This field is termed the matching field, which for the substrates used in this studies is $H_m=0.22$ T.  In fact, any integer multiple of $H_m$ will also lead to a cancellation of the screening currents.  In our system, however, the critical field is reached before a second dip or reentrance is observed.

\begin{figure}
\begin{flushleft}
\includegraphics[width=.44\textwidth]{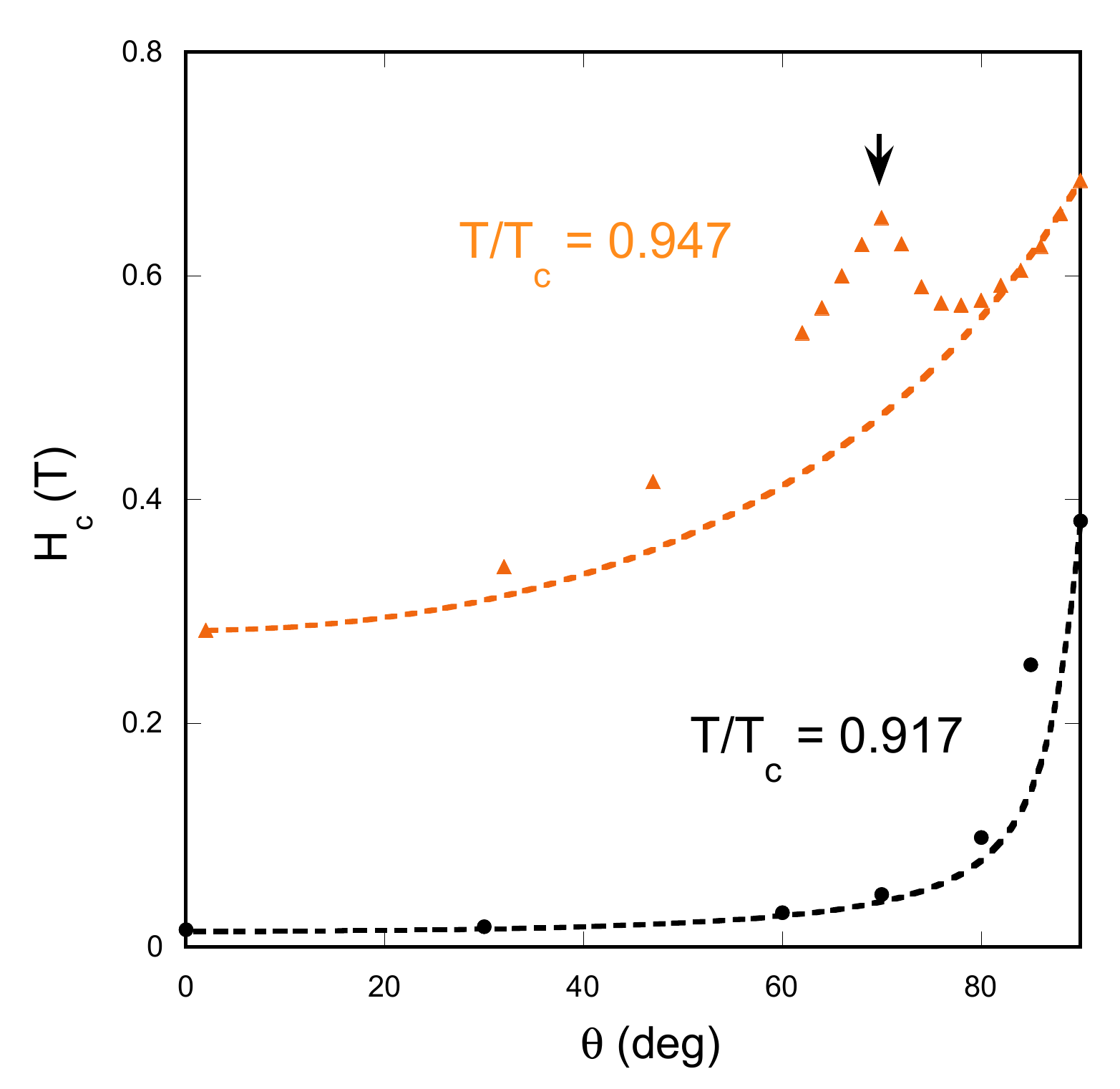}\end{flushleft}
\caption{Black symbols: midpoint critical field values for a 9 nm film on glass (see data in Fig.~\ref{R-H-uniform}) as a function of tilt angle.  The dashed line represents the predicted angular dependence of the Tinkham formula, Eq.~(\ref{2D}). Orange symbols: midpoint critical field values for a 9 nm Al film on an AAO substrate (see data in Fig.~\ref{R-H-9}).  The dashed line represents the predicted angular dependence of the Tinkham formula.}
\label{Hc-angle}
\end{figure}

Shown in Fig.~\ref{Hc-angle} as black symbols are the midpoint critical fields from Figs.~\ref{R-H-uniform} as a function of tilt angle at a temperature relatively close to $T_c$.  The dashed line represents the angular dependence as predicted by the Tinkham formula \cite{2D-Tinkham},

\be
\frac{H_c(\theta)\cos(\theta)}{H_{c2}}+\left(\frac{H_c(\theta)\sin(\theta)}{H_{c\parallel}}\right)^2=1.
\label{2D}
\ee

For comparison we have also plotted as red symbols the midpoint critical fields obtained from Fig.~\ref{R-H-9}.  We point out  that Eq.~(\ref{2D}) has no adjustable parameters.  Note the excellent agreement between Eq.~(\ref{2D}) and the angular dependence of the uniform film critical field.   In contrast, critical fields of the AAO film tend to be somewhat higher than the Tinkham curve and exhibit a local maximum as indicated by the arrow.  Although the angular dependence of the critical field data can be sensitive to the critical field criterion, the AAO data will, nevertheless, not fall on the Tinkham curve regardless of which criterion is used due to the presence of the local maximum.

\section{Discussion}

There are three primary effects associated with rotating out of perpendicular orientation.  The first is an increase in the critical field.  This occurs because the orbital pair-breaking energy is proportional to the square of the characteristic transverse dimension.  In perpendicular field the transverse dimension is the superconducting link width $\sim30\,$nm and in parallel field it is the film  thickness, which is 9~nm.  The second is that as a consequence of a higher critical field the role of Zeeman splitting, or spin polarization, becomes more important in the transition region.  The third effect is to push the matching field reentrance feature, denoted by $H^*$, to higher and higher fields until, in principle, one reaches a condition where the matching-field resistance minimum occurs at the global critical field, {\it i.e.} $H^*\sim H_c$.

\begin{figure}
\begin{flushleft}
\includegraphics[width=.44\textwidth]{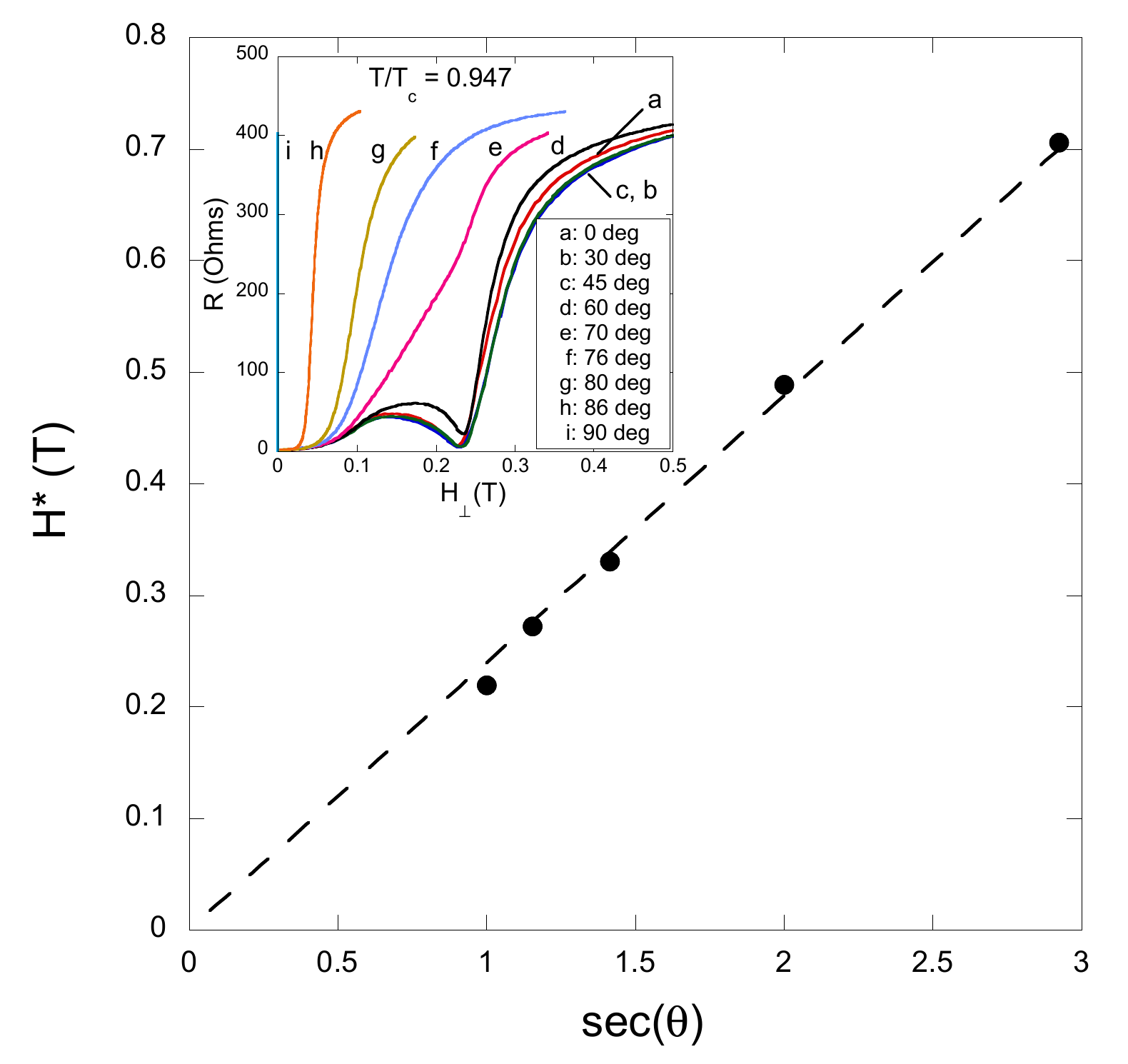}
\end{flushleft}
\caption{ The matching fields from the data in Fig.~\ref{R-H-9} as a function of $\sec(\theta)$.  The dashed line represents a linear least-squares fit to the data.  The slope of the fit corresponds to a matching field of $H_m=0.24$ T. Inset: critical field transitions of Fig.~\ref{R-H-9} plotted as a function of the perpendicular component of the magnetic field.}
\label{H-match}
\end{figure}

The uniform Al film exhibits a monotonic increase in the measured critical field, regardless of which commonly used criterion is applied to define the critical field {\textemdash} 1\%, 10\%, or 50\% of the normal state resistance $R_n$, see Fig.~\ref{R-H-uniform}.  This is clearly not case for the nano-pore film, as can be seen in Fig.~\ref{R-H-9}.  Indeed, the presence of a shifting and broadening reentrance minimum in the resistivity traces makes the definition of critical field somewhat ambiguous.  Notwithstanding this issue, the resistivity minimum should occur when the perpendicular component of the field is equal to the matching field, thus the matching field reentrance feature occurs at $H^*=H_m\sec(\theta)$.  Shown in Fig.~\ref{H-match} is a plot of the position of the resistivity minimum as a function of $\sec(\theta)$.  Note the expected linear dependence, which suggests that the evolution of the resistivity traces with increasing angle in Fig.~\ref{R-H-9} is a consequence of phase effects in the superconducting network.  Of course, the global critical field $H_c$ also increases with increasing angle as per Eq.~(\ref{2D}).  However, $H_c$ increases more slowly than $H^*$ due to the finite $H_{c\parallel}$. In fact, the R-H traces up to and including the 60$^\circ$ curve appear to be similar to each other, in the sense that there is a well defined dissipation minimum when $H\cos\theta=H_m$.  In contrast, not only does the 70$^\circ$ trace not exhibit a local minimum, but its width extends across the entire critical region.  At this angle the matching field is comparable to the critical field, $H^*\sim H_c=0.7$ T and, indeed, there is a barely discernible inflection point in the trace at 0.7 T, see inset of Fig.~\ref{H-match}.  Furthermore, the dissipation peak that appears at $H\approx H^*/2$ in the $\theta=0^\circ$ trace of Fig.~\ref{R-H-9}, is broadened by a factor of 3 at $\theta=70^\circ$.  This sub-matching field dissipation appears to dominate the transition region at this angle.  At higher angles the $H^*$ is pushed well beyond the upper critical field and the transitions look somewhat more conventional.

In principle, one should be able to use the matching field reentrance to enhance the parallel critical field of a thin film superconductor.  If one neglects orbital pair-breaking effects of the parallel component of the applied field, which is reasonable for Al films of thickness less than 4 nm, then the $T=0$ parallel critical field is Zeeman limited and given by the Clogston-Chandrasekhar equation \cite{CC} $\displaystyle H_{c\parallel}=\frac{\sqrt2\Delta_0}{g\mu_B}$, where $\Delta_0$ is the zero temperature - zero field gap energy.  Under these conditions, the Zeeman critical field represents the maximum possible critical field of a homogenous superconductor.  Any rotation off of parallel results in a finite perpendicular component of the field which, by way of orbital pair-breaking, lowers the critical field.   In contrast, the role of orbital pair-breaking effects in a multiply connected superconductor network are not as straightforward due to flux quantization.  In particular, at the matching field the screening currents of the network cancel and the system can reenter a dissipation-less phase.    Therefore, the maximum critical field is not obtained at parallel orientation but, instead, slightly off of parallel so that the perpendicular component of the field is equal to $H_m$.  Under these conditions, the critical field transition is still driven by the Zeeman splitting but the applied field is now larger $H^{max}_c=\sqrt{{H_{c\parallel}}^2+{H_m}^2}$. For a triangular array of pores with lattice constant $a$ the matching field is given by $\displaystyle H_m=\frac{2\Phi_0}{\sqrt3a^2}$, where $\Phi_0$ is the flux quantum \cite{Valles1}. Assuming $H_{c\parallel} \gg H_m$ and taking $g=2$, the resulting enhancement to the Zeeman-limited critical field is,

\be
H^{max}_c\approx H_{c\parallel}\left(1+\frac{4{\Phi_0}^2{\mu_B}^2}{3a^4{\Delta_0}^2}\right)
\label{HcMax}
\ee

Although we observed a critical field enhancement consistent with Eq.~\ref{HcMax} in many samples, the AAO substrates where simply not sufficiently flat and smooth to consistently map out the critical field behavior in the angular region near $\theta=90^\circ$ \cite{Valles3}.  However, a local maximum in the angular dependent critical field, such as the peak highlighted by the arrow at 70$^\circ$ in Fig.~\ref{Hc-angle} was often seen.  This particular peak occurs at a critical field of $H_c=0.653$ T.  The corresponding perpendicular component of the field is
equal to the matching field $H_\perp=0.653\times\cos70^\circ=0.22$ T, as expected.

\begin{figure}
\begin{flushleft}
\includegraphics[width=.44\textwidth]{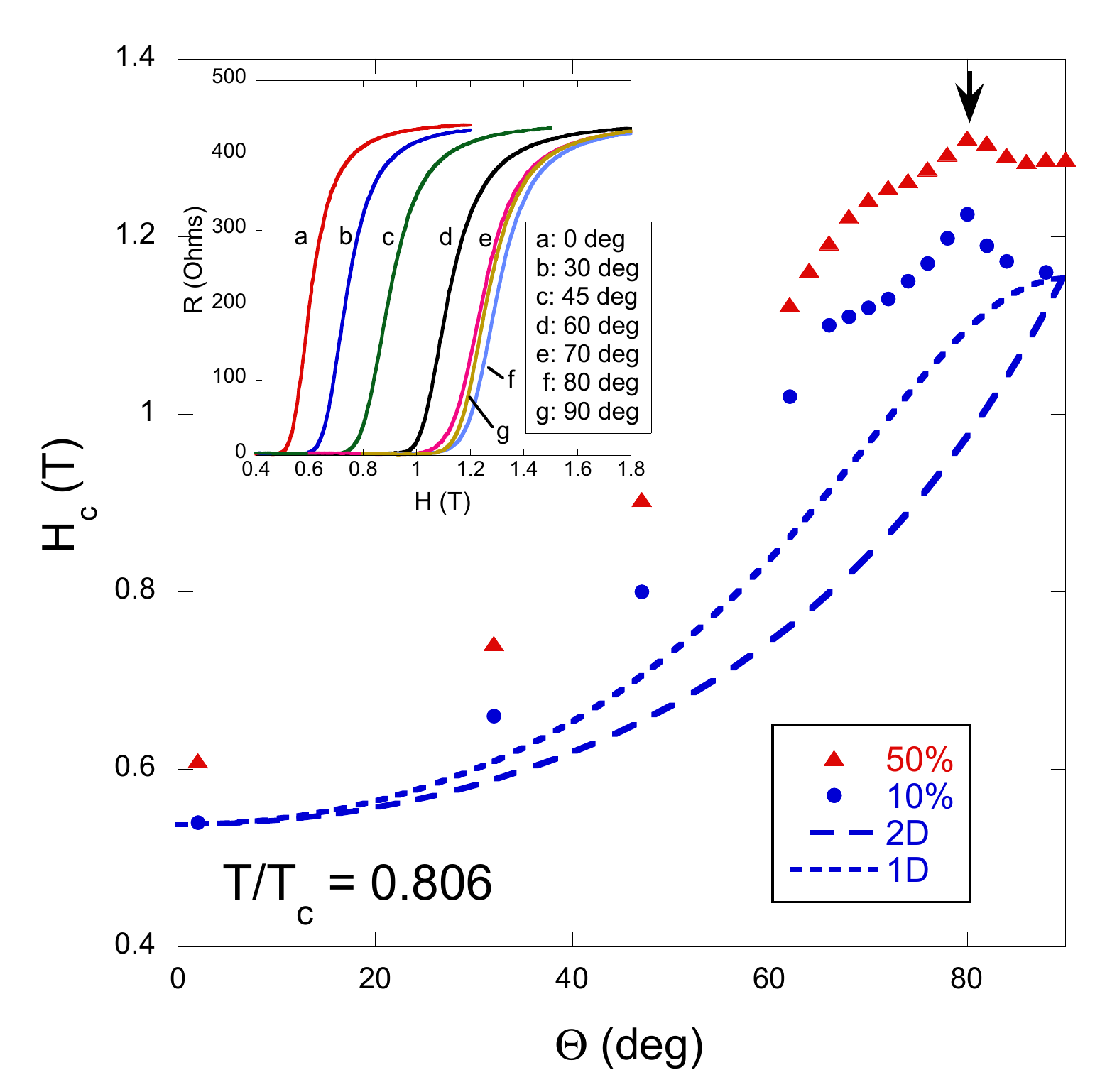}
\end{flushleft}
\caption{Red symbols: midpoint critical field values for the 9 nm film on AAO used in Fig.\ \ref{R-H-9}.  These data were taken at 1.83 K.  Blue symbols: critical field values defined by the field at which the resistance is 10\% of the normal state resistance.  The dashed lines represent the predicted angular dependence of the 1D and 2D Tinkham formulas, see text. Inset: Resistive critical field transitions of the AAO sample from Fig.\ \ref{R-H-9} taken at $T=1.83$ K.  The reentrant features are no longer present at this lower temperature.}
\label{Hc-angle-LT}
\end{figure}

Determining the critical field from reentrant $R-H$ traces, such as the ones in Fig.\ \ref{R-H-9} is problematic
due to the fact that if too low of a critical field criterion is used, such as when the resistance reaches 10\% of its normal state value, then one can have multiple crossings of the threshold at a given angle.  However, at lower temperatures the dissipation associated with sub-matching fields are too small to be observed over a more robust superconducting condensate.  Consequently the $R-H$ traces are more conventional as can be see in the inset of Fig.\ \ref{Hc-angle-LT}.  These data are from the same sample as Fig.\ \ref{R-H-9} but were taken at $T=1.83$ K.  The corresponding critical fields, defined by both a midpoint and a 10\% criterion, are shown in the main panel of the figure.  The dashed lines represent the expected behavior of the thin film (2D) Tinkham formula from Eq.~\ref{2D} and the corresponding 1-dimensional version of the formula \cite{1D-Tinkham}.  The latter assumes the superconducting necks between the holes can be modeled as a collection superconducting slabs having width 30 nm and thickness 9 nm and that the coherence length is longer than either dimension. Note that neither model predicts the observed angular dependence.  This may, in part, be a consequence of the Zeeman contribution to the critical field.  We point out that the Zeeman response is not a significant factor in behavior of the Nb films used in Refs. \cite{Patel} and \cite{1D-Tinkham} due to the large spin-orbit scattering rate of Nb.

The critical fields in Fig.~\ref{Hc-angle-LT} are about a factor of 2 higher than their higher temperature counterparts shown in Fig.~\ref{Hc-angle}.  Consequently one would expect the local maximum in $H_c$ to shift to a larger angle.  Indeed, as indicated by the arrow in Fig.~\ref{Hc-angle-LT} a maximum critical field of $Hc\sim1.3$ T is obtained at $80^\circ$ with a corresponding perpendicular component $H_\perp=1.3\times\cos80^\circ=0.22$ T, that is near the matching field.  Interestingly, although the critical field traces in the inset of Fig.~\ref{Hc-angle-LT} show no signs of reentrance at the matching field, the effects of flux quantization are nevertheless manifest in the off-parallel maximum indicated by the arrow.

In summary, we have measured the critical field of Al films deposited on AAO nano-pore substrates as a function of tilt angle.  We find that near $T_c$ the angle-dependent critical field behavior of these multiply connected superconducting films is drastically different than that of a uniform film of similar thickness.   In particular, when the tilt angle is set so that $H^*\sim H_c$ the transition becomes extremely broad, suggesting that the dissipation associated with a sub-matching field dominates the critical behavior.  Flux quantization can also produce an off-parallel maximum in the critical field when the perpendicular component of the magnetic field equals the matching field.

\acknowledgments

The magneto-transport measurements were performed by P.W.A. and F.N.W with the support of the U.S. Department of Energy, Office of Science, Basic Energy Sciences, under Award No.\ DE-FG02-07ER46420.    The theoretical analysis was carried out by G.C. The AAO substrates were provided by Jimmy Xu, Chunshu Wu, and Jin Ho Kim.

\end{document}